\documentclass[aps,prl,twocolumn,superscriptaddress]{revtex4}

\usepackage[dvips]{graphicx}
\usepackage{times}

\newcommand{\br}{\mbox{\boldmath $r$}}

\newcommand{\bk}{\mbox{\boldmath $k$}}

\begin{document}

\title{First-principles Calculation of 
       Effective Onsite Coulomb Interactions
       of 3$d$ Transition Metals:
       Constrained Local Density Functional Approach
       with Maximally Localized Wannier Function}

\author{Kazuma Nakamura} 
\thanks{Electronic mail: kazuma@cms.phys.s.u-tokyo.ac.jp}
\affiliation{Department of Physics, Graduate School of Science, 
 University of Tokyo, 7-3-1 Hongo, Bunkyo-ku, Tokyo 133-0033, Japan}
 
\author{Ryotaro Arita} 
\thanks{Present address: Max-Planck-institut f\"{u}r Festk\"{o}rperforschung,
 Heisenbergstrasse 1, D-70569 Stuttgart, Germany.} 
\affiliation{Department of Physics, Graduate School of Science, 
 University of Tokyo, 7-3-1 Hongo, Bunkyo-ku, Tokyo 133-0033, Japan}

\author{Yoshihide Yoshimoto} 
\affiliation{Institute for Solid State Physics, 
 University of Tokyo, 5-1-5 Kashiwanoha, Kashiwa, Chiba 277-8531, Japan} 

\author{Shinji Tsuneyuki}
\affiliation{Department of Physics, Graduate School of Science, 
 University of Tokyo, 7-3-1 Hongo, Bunkyo-ku, Tokyo 133-0033, Japan} 

\date{\today}

\begin{abstract}
 We present a new {\em ab initio} method 
 for calculating effective onsite 
 Coulomb interactions of itinerant
 and strongly correlated electron systems. 
 The method is based on constrained local density functional theory 
 formulated in terms of 
 maximally localized Wannier functions. 
 This scheme can be implemented with any basis,  
 and thus allows us to perform 
 the constrained calculation 
 with plane-wave-based electronic-structure codes.  
 We apply the developed method to 
 the evaluation of the onsite interaction of
 3$d$ transition-metal series.      
 The results are discussed using           
 a heuristic formula for screened Coulomb interactions.     
\end{abstract}

\pacs{}

\maketitle
 
 Properties of itinerant and strongly correlated
 electron systems are widely discussed with 
 phenomenological Hamiltonians such as
 Anderson-impurity 
 \cite{Ref_Anderson}
 and Hubbard models 
\cite{Ref_Hubbard}.             
 These models embody essential aspects
 of correlated electrons,   
 and can be solved numerically or
 sometimes analytically for special cases.        
 However, parameters in the models
 are often determined empirically so that the employed model
 should reproduce experimental results of interest.         

 Recently, there has been growing interest in constructing model 
 Hamiltonians from first principles \cite{Ref_WF}. 
 The principal motivation for such a study   
 is a so-called long-standing ``beyond LDA'' problem. 
 One of recent major approaches 
 toward correlated electrons from first principles is combining 
 density-functional theory within
 local density approximation (LDA) with dynamical 
 mean-field theory (DMFT) \cite{Ref_LDA+DMFT}.
 The main idea of this method is to map 
 a first-principles Hamiltonian  
 onto a lattice-fermion model to which the DMFT method is applicable,
 so developments of tractable but rational mapping
 techniques are highly needed.    

 An important notice in constructing {\em ab initio} model Hamiltonians  
 lies in an evaluation of an onsite
 interaction parameter known as the Hubbard $U$. 
 Transfer parameters can be easily obtained by representing 
 a one-body {\em ab initio} Hamiltonian  
 with a spatially localized (or atomic) orbital, 
 while the $U$ parameter evaluated with
 a one-center Coulomb integral of the atomic orbital 
 gives a rather large value relative to 
 that deduced from 
 experiments \cite{Ref_Sawatzky,Ref_Yin}. 
 This is because a screening effect     
 of surrounding valence electrons 
 is completely neglected in the evaluation of the value.  
 A representative treatment for calculating $U$ 
 including the screening effect
 is a constrained approach \cite{Ref_CLDA_1, Ref_GW}.      
 It gives an optimally screened $U$, because, in that calculation, 
 the $U$ value is obtained from a response of the system 
 to a change in a local charge density, thus  
 incorporating the valence-electron-screening
 effect statically into the $U$ calculation. 

 An important point in performing the constrained calculation 
 is a choice of basis functions 
 to define the charge density of localized electrons. 
 The use of an atom-centered localized basis set
 like a linear muffin-tin orbital (LMTO) \cite{Ref_LMTO}           
 facilitates this definition,        
 so constrained calculations
 to date have been performed
 with LMTO-based electronic-structure codes.  
 The method has been widely applied to investigations   
 of effective Hubbard $U$'s of
 various correlated materials \cite{Ref_CLDA_2, Ref_Anisimov}, 
 while it is known that the calculated effective onsite
 Coulomb interaction includes an error \cite{Ref_Anisimov} arising from       
 the atomic sphere approximation employed in 
 the conventional LMTO calculations \cite{Ref_NMTO}. 

 In this Letter, we present a new method for calculating an effective $U$;
 that is, a constrained local density functional approach 
 based on ``maximally localized'' Wannier functions (WF's) \cite{Ref_MLWF}. 
 The maximally localized WF, obtained by minimizing a spatial spread 
 of WF's in real space,  
 has a practical advantage that it can be computed 
 with any basis functions. 
 This property enables us to implement the constrained scheme 
 in the plane-wave-based electronic-structure codes. 
 In addition, the maximally localized WF is regarded as a reasonable 
 basis for constructing model Hamiltonians; 
 the property of the WF being well-localized spatially  
 ensures that the Hamiltonian matrix represented by this basis 
 is sparse and short-ranged. 
 In the present study, we apply the developed method
 to the systematic analysis 
 for effective onsite Coulomb interactions      
 of 3$d$ transition metals \cite{Ref_Schnell}. 

 An electronic structure of a transition metal
 consists of localized $d$ electrons forming 
 a narrow band and itinerant $sp$ electrons
 associated with a wider band.         
 There are $n$ localized electrons per atom for the ground state.     
 An effective Coulomb interaction $U_{{\rm eff}}$
 between two $d$ electrons in an atom
 is defined as an energy cost in 
 the electron transfer process, $2 d^n \to d^{n-1} + d^{n+1}$,  
 which is written in terms of total energies as 
\begin{eqnarray}
 U_{{\rm eff}} = E(n+1) + E(n-1) - 2 E(n), \label{eq:Ueff}  
\end{eqnarray}
 where $E(n)$ is the ground-state energy of the system, while  
 $E(n\pm1)$ correspond to total energies          
 for cases where there are $n\pm1$ localized electrons
 in a specific atom. 
 It should be noted here that the number of onsite $d$ electrons 
 is regarded as an adiabatic parameter 
 for the electronic state of the system; i.e., 
 for the total-energy calculation, the charge density of the system
 is relaxed in the constraint 
 that the occupation number of localized $d$ orbitals  
 of the specific atom is kept at a given value.  
 The $U_{{\rm eff}}$ value 
 thus obtained 
 includes a screening effect
 due to the relaxation of the valence-electron
 density around the specific atom.         

 The total energies can be basically calculated
 in density-functional formalism \cite{Ref_DFT}, 
 but an important point is 
 how to impose the constraint mentioned above.  
 To this end, we first define
 the $d$ occupation number of the atom $I$ as 
\begin{eqnarray}
 N_{Id} = \sum_{\mu}^{d} \sum_{\bk}^{{\rm BZ}} \sum_{\alpha}^{occ} 
          f_{\alpha \bk} \bigl| \langle w_{I\mu} |
          \phi_{\alpha \bk} \rangle \bigr|^2, \label{eq:NId} 
\end{eqnarray}
 where $\phi_{\alpha \bk}$ is 
 a Bloch orbital of a band        
 $\alpha$ with a wave vector $\bk$, 
 $f_{\alpha \bk}$ is its occupation number, 
 \{$w_{I\mu}$\} are $d$-type
 maximally localized Wannier functions centered at the atom $I$,
 and the index $\mu$ specifies the five types of the $d$ orbital. 
 With the definition of the $N_{Id}$ above,
 we may write a constrained density functional as 
\begin{eqnarray}
 E(N_{Id}) &=& \min_{\rho} \Biggl\{ F[\rho (\br)]   
 - \lambda \biggl[ \sum_{\mu}^{d} \sum_{\bk}^{{\rm BZ}}
     \sum_{\alpha}^{occ}   \nonumber \\ 
 & & f_{\alpha \bk} \bigl| \langle w_{I\mu} | \phi_{\alpha \bk} \rangle \bigr|^2 
 - N_{Id} \biggr] \Biggr\}. \label{eq:CDF} 
\end{eqnarray}
 Here, $F[\rho]$ is a usual density functional
 with a total charge density $\rho(\br)=\sum_{\alpha \bk}|\phi_{\alpha \bk}(\br)|^2$, and  
 $\lambda$ is a Lagrange multiplier. 
 A functional derivative of $E(N_{Id})$ with respect to
 the Bloch orbital $\phi_{\alpha\bk}$ 
 leads to the following constrained Kohn-Sham (KS) equation, 
\begin{eqnarray}
 \biggl[ H_{KS} 
  + \lambda \sum_{\mu}^{d} | w_{I\mu} \rangle \langle w_{I\mu} | \biggr]
  \phi_{\alpha \bk} =
 \epsilon_{\alpha \bk} \phi_{\alpha \bk},   \label{eq:CKS}  
\end{eqnarray}
 where $H_{KS}$ is a one-body KS Hamiltonian, and 
 the second term in the brackets is an
 additional potential due to the constraint. 

 The calculation for the effective onsite
 Coulomb interaction $U_{{\rm eff}}$ proceeds as follows:
 We first perform a total-energy density-functional calculation
 with no constraint ($\lambda=0$) to obtain the ground-state energy $E(n)$. 
 Then, we transform the resulting Bloch functions \{$\phi_{\alpha \bk}^{\lambda=0}$\}
 into the maximally localized WF's 
 \{$w_{I\mu}^{\lambda=0}$\} used as
 the input of the subsequent constrained calculations.
 We then solve the constrained KS equation
 [Eq.~(\ref{eq:CKS})] with a non-zero $\lambda$, 
 and calculate the occupation number $N_{Id}$ in Eq.~(\ref{eq:NId})
 using the resulting Bloch functions \{$\phi_{\alpha \bk}^{\lambda \ne 0}$\}
 and the WF's above \{$w_{I\mu}^{\lambda=0}$\}.  
 The $\lambda$ parameter is adjusted so that $N_{Id}$ 
 should be a desired number; i.e., $n-1$ or $n+1$.         
 Finally, we calculate $U_{{\rm eff}}$
 of Eq.~(\ref{eq:Ueff})  
 with the total energies thus obtained, $E(n)$, $E(n-1)$, and $E(n+1)$. 

 We implemented the scheme presented here in 
 {\em Tokyo Ab initio Program Package} \cite{Ref_TAPP} developed 
 by the condensed-matter-theory group
 in the University of Tokyo.  
 With this program, band calculations 
 were performed within the generalized gradient 
 approximation \cite{Ref_PBE96} to density-functional theory, 
 using a plane-wave basis set
 and the Troullier-Martins
 norm-conserving pseudopotentials \cite{Ref_PP1} 
 in the Kleinman-Bylanger representation \cite{Ref_PP2}. 
 The energy cutoff was set to 64 Ry, 
 and a 9$\times$9$\times$9 $k$-point sampling 
 was employed to represent electronic
 structures of transition metals \cite{Ref_Kittel}. 
 A spin polarization effect was neglected
 in the present calculations,  
 and this effect seems to be negligible 
 for the evaluation of $U_{{\rm eff}}$ \cite{Ref_Anisimov}. 
 The $d$-type WF's were constructed 
 with the algorithm of Souza, Marzari, and Vanderbilt \cite{Ref_MLWF}.  
 To show the validity of representing
 $d$ bands by the resulting WF's,  
 we compare in Fig.~1 original bands of Fe (solid line)
 with interpolated bands (dotted line) 
 obtained by diagonalizing $k$-space Hamiltonian matrix represented by 
 the five $d$-type WF's.  
 We see that the calculated interpolated bands reproduce 
 reasonably the original narrow $d$ bands,          
 except when $sp$ bands cross the $d$ bands. 
 Constrained calculations were performed for
 a suppercell containing 8 transition-metal atoms 
 to consider a full relaxation of a screening charge density
 to a local charge perturbation. 
 It was found that the size of the suppercell
 is sufficient for obtaining a converged $U_{{\rm eff}}$ value.
\begin{figure}[h]
\begin{center}
\includegraphics[width=6.5cm]{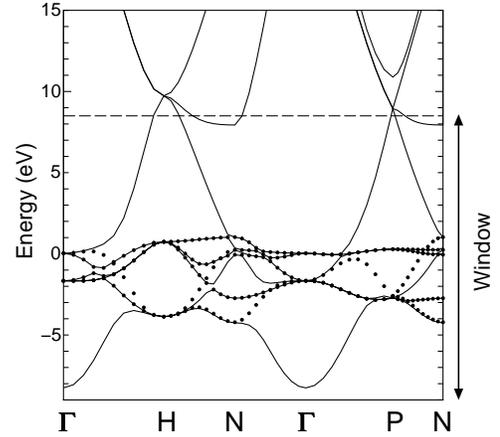}
\end{center}
\caption{Solid line: Calculated band structure of Fe.
         Dotted line: Interpolated bands obtained from the five 
         $d$-type Wannier functions. Energy window \cite{Ref_MLWF} used to compute 
         the Wannier functions was set to [$-$9.0,$+$8.5] eV. 
         The zero of the energy scale is at the Fermi energy.}
\end{figure}
     
 We show in Fig.~2 the calculated effective
 onsite Coulomb interaction $U_{{\rm eff}}$
 (solid line) of the 3$d$ metals. 
 An importance of a screening effect is clear 
 from a comparison between the $U_{{\rm eff}}$ value and 
 a bare Coulomb integral averaged over five $d$-type WF's,  
\begin{figure}[h]
\begin{center}
\includegraphics[width=5.5cm]{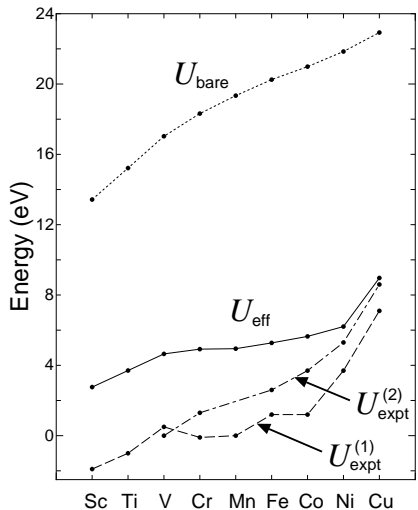}
\end{center}
\caption{$U_{{\rm eff}}$ (solid line), 
         $U_{{\rm bare}}$ (dotted line),
         and two experimental onsite
         Coulomb interactions (dashed and dotted-dashed lines).  
         The experimental values are taken from
         Ref.~\cite{Ref_Sawatzky} for $U_{{\rm expt}}^{(1)}$
     and Ref.~\cite{Ref_Yin}      for $U_{{\rm expt}}^{(2)}$.}
\end{figure}
\begin{eqnarray}
 U_{{\rm bare}} = \frac{1}{5} \sum_{\mu}^{d}
 \int \int \frac{|w_{I\mu}(\br)|^2 |w_{I\mu}(\br')|^2}
 {|\br - \br'|} d\br d\br'. \label{eq:Ubare}  
\end{eqnarray}
 As expected, an unscreened value $U_{{\rm bare}}$ (dotted line) 
 is rather large relative to 
 $U_{{\rm eff}}$. 
 The screening effect 
 makes a significant reduction of the value, 
 thus leading the theoretical values to the same energy order as 
 experimental values (dashed \cite{Ref_Sawatzky}
 and dotted-dashed \cite{Ref_Yin} lines)              
 deduced from a combined 
 use of Auger and x-ray photoemission spectroscopy.   
 A more detailed comparison with the $U_{{\rm expt}}$'s       
 requires to calculate {\em ab initio}
 Auger and photoemission spectra,
 but this is not the scope of the present study.   

 An interesting behavior seen in the figure is that 
 $U_{{\rm eff}}$ increases gradually 
 as an atomic number increases, 
 and this trend agrees with the experiments.  
 To our knowledge, there is no explanation 
 for the chemical trend observed in $U_{{\rm eff}}$, 
 so we focus on understanding this behavior. 
 If the screening effect is neglected,               
 $U_{{\rm eff}}$ is $U_{{\rm bare}}$,   
 for which it is apparent from Eq.~(\ref{eq:Ubare})
 that the value of $U_{{\rm bare}}$ 
 reflects a feature of the WF itself. 
 More specifically, 
 the observed increasing behavior of $U_{{\rm bare}}$
 results from the trend that        
 the WF shrinks as an early metal goes to a late one, 
 which will be also stated in terms of the Wannier spread;  
 namely, $U_{{\rm bare}}$ 
 is inversely scaled by a localization length
 describing a spatial extent of WF 
\begin{eqnarray}
 \Omega = \frac{1}{5} \sum_{\mu}^{d}
 \sqrt{\langle r^2 \rangle_{\mu} -
 \langle \br \rangle_{\mu}^{2} },   
  \label{eq:OMEGA}  
\end{eqnarray}
 where $\br$ is electron coordinates and 
 $\langle \br \rangle_{\mu} = \int \br |w_{I\mu}(\br)|^2 d\br$, etc.  

 In contrast to $U_{{\rm bare}}$,
 the analysis for $U_{{\rm eff}}$ itself
 is rather intractable, 
 because, in that case, in addition to the shrinking effect above,       
 a difference in the screening in the metals  
 also contributes to the trend in $U_{{\rm eff}}$. 
 In the transition metal, the screening 
 results mainly from itinerant $sp$ electrons 
 behaving as a free electron,  
 so the effective Coulomb
 interaction between two $d$ electrons
 is to a first approximation written
 as an interaction between test charges placed 
 in a homogeneous electron gas, i.e.,            
 as the Yukawa-type interaction  $e^{-(r/\sigma_{0})}/r$  
 with $\sigma_{0}$ being the Thomas-Fermi screening length \cite{Ref_Ziman}. 
 From this viewpoint, 
 we give a heuristic formula for $U_{{\rm eff}}$,   
 which is assumed to be a one-center
 integral of the Yukawa potential as \cite{Ref_Norman}  
\begin{eqnarray}
 \tilde{U}_{{\rm eff}} &=& \frac{1}{5} \sum_{\mu}^{d}
 \int \int \frac{|w_{I\mu}(\br)|^2 \exp (- |\br - \br'| / \sigma )} 
 {|\br - \br'|} \nonumber \\ 
 & & \times |w_{I\mu}(\br')|^2 
  d\br d\br', \label{eq:Uscr}  
\end{eqnarray}
 where $\sigma$ in the integral is an adjustable parameter to 
 specify a characteristic length of the screening in the system.         
 As the value of $\sigma$ becomes larger (i.e., the screening 
 is weakened), $\tilde{U}_{{\rm eff}}$
 approaches to          
 $U_{{\rm bare}}$                       
 (unscreened value),  
 so this parameter is regarded as a measure of the screening strength.     
 We determined the $\sigma$ parameter 
 so that the integral value $\tilde{U}_{{\rm eff}}$ should coincide 
 with the $U_{{\rm eff}}$ obtained
 from {\em ab initio} calculations.  
 By plotting $\sigma$ for all the 3$d$ species, 
 we can discuss the trend in
 the screening strength of the transition metals.       

 We display in Fig.~3 the resulting dependence of $\sigma$
 on atomic species (dashed line), 
 together with that of $\Omega$ (dotted line). 
 The figure clearly shows that 
 the increasing behavior 
 of $U_{{\rm eff}}$ (solid line) results from 
 the orbital shrink
 (i.e., the decrease of $\Omega$).  
 In contrast,
 the effective screening length $\sigma$ is almost the same 
 for the listed metals, 
 although there is a small jump of $\sigma$ between Ni and Cu,  
 which reflects the fact that Ni has
 two screening channels of $sp$ and $d$ electrons, 
 while Cu has no latter channel
 because the Cu $d$ bands are fully occupied. 
 We note that the calculated $\sigma$ agrees reasonably well 
 with the Thomas-Fermi screening length $\sigma_{0}$ (dotted-dashed line) 
 of a free electron gas with the same density as the 
 $sp$-electron density of the system \cite{Ref_Nsp}, 
 which indicates a validity of our
 $\sigma$ estimation based on Eq.~(\ref{eq:Uscr}).                     
 The small difference between
 $\sigma$ and $\sigma_{0}$ ($\sim$ 0.1 \AA)
 reflects the presence or absence of a screening contribution 
 from intersite $d$ electron transfers.  
\begin{figure}[h]
\begin{center}
\includegraphics[width=6.0cm]{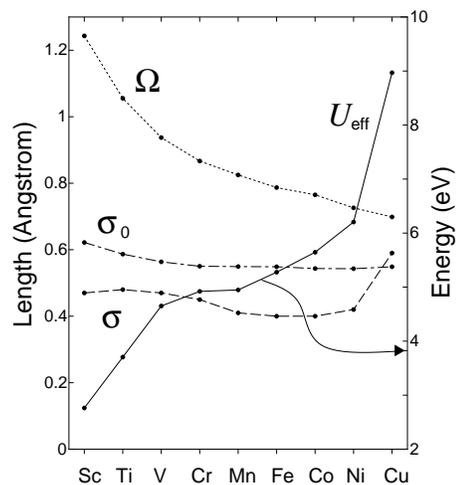}
\end{center}
\caption{$U_{{\rm eff}}$ (solid line), 
         Wannier spread $\Omega$ (dotted line),
         effective screening length $\sigma$ (dashed line), and                  
         Thomas-Fermi screening length $\sigma_{0}$ (dotted-dashed line).} 
\end{figure}

 In summary, we have developed
 a constrained local density functional 
 approach with maximally localized Wannier functions, and 
 applied it to the systematic study of the effective 
 onsite Coulomb interaction of 3$d$ transition metals. 
 This scheme is of practical significance because 
 it is free from a limitation on a choice of basis functions, 
 and thus allows us to perform the
 constrained calculations with plane-wave-based 
 electronic-structure codes. 
 This property will be particularly helpful in a situation where 
 one's interest is in molecular solids
 such as BEDT-TTF \cite{Ref_BEDT-TTF}
 and solid C$_{60}$ \cite{Ref_C60_1,Ref_C60_2},       
 for which the plane-wave basis is suitable for calculating         
 their electronic structures.  
 To see whether the present scheme can be indeed exploited 
 in constructing {\em ab initio} model Hamiltonians
 and/or in discussing experimental $U$'s          
 for the above mentioned systems,            
 however, needs future studies.  

 This work was supported by NAREGI Nanoscience Project, 
 Ministry of Education, Culture, Sports, Science and 
 Technology, Japan. 
 One of us (K.N.) acknowledges Research Fellowships of 
 the Japan Society for the Promotion of 
 Science for Young Scientists.              
 All the calculations were performed 
 on Hitachi SR11000 system 
 of the Super Computer Center at the Institute for 
 Solid State Physics, the University of Tokyo.          
 The authors would like to thank Kazuto Akagi 
 for his advice on the program development, 
 and Atsushi Yamasaki for useful discussions.


\begin{references}
\bibitem{Ref_Nsp}
 The number of $sp$ electrons $N_{sp}$ is defined as 
 $N_{sp} = N - N_{d}$, where $N$ is the total number 
 of the electrons of the system, while $N_{d}$ is 
 the number of $d$ electrons defined via summing up 
 the onsite occupation number $N_{Id}$ of Eq.~(\ref{eq:NId})        
 over sites \{$I$\} in a cell.
\end{references}
\end{document}